\documentclass[prb,twocolumn,english,showkeys,showpacs,amsmath,amsfonts,floatfix,superscriptaddress,nofootinbib]{revtex4-1}

\def\be{\begin{equation}}
\def\ee{\end{equation}}
\def\bea{\begin{eqnarray}}          
\def\eea{\end{eqnarray}}
\def\bi{\begin{itemize}}
\def\ei{\end{itemize}}
\def\nb{\nonumber}
\usepackage{xcolor}
\usepackage{graphicx}
\usepackage{amsmath,amstext,amssymb,babel,bm,color,times}

\newcommand{\sign}{\text{sgn}}
\usepackage[colorlinks=true]{hyperref}

\begin{document}

\title{ 
         Is there a correlation length in 
         a model with long-range interactions?
}

\author{Debasis Sadhukhan} 
\author{Jacek Dziarmaga} 
\affiliation{Jagiellonian University, Institute of Theoretical Physics, 
             {\L}ojasiewicza 11, PL-30348 Krak\'ow, Poland}

\date{ July 5, 2021 }


\begin{abstract}
Considering an example of the long-range Kitaev model, we are looking for a correlation length in a model with long range interactions whose correlation functions away from a critical point have power law tails instead of the usual exponential decay. 
It turns out that quasiparticle spectrum depends on a distance from the critical point in a way that allows to identify the standard correlation length exponent, $\nu$. The exponent implicitly defines a correlation length $\xi$ that diverges when the critical point is approached. 
We show that the correlation length manifests itself also in the correlation function but not in its exponential tail because there is none. Instead $\xi$ is a distance that marks a crossover between two different algebraic decays with different exponents. At distances shorter than $\xi$ the correlator decays with the same power law as at the critical point while at distances longer than $\xi$ it decays faster, with a steeper power law. For this correlator it is possible to formulate the usual scaling hypothesis with $\xi$ playing the role of the scaling distance. 
The correlation length also leaves its mark on the subleading anomalous fermionic correlator but, interestingly, there is a regime of long range interactions where its short distance critical power-law decay is steeper than its long distance power law tail.  
\end{abstract}
\maketitle


{\bf Long range (LR) interactions. ---}
LR interactions are common in nature. Examples spread from self-gravitating systems\cite{Padmanabhan, Chavanis}, through dipolar ferromagnets \cite{LandauLifshitz,dipolExp1, dipolExp2}, to spin-ice materials \cite{spinice1, spinice2}. Moreover, over the last few decades tremendous advancements in cold-atomic techniques have established our ability to control and manipulate LR interacting models at an unprecedented level ~\cite{iontrapexp1,Friedenauer2008,PhysRevLett.103.120502,Kim2010,PhysRevB.82.060412,Lanyon57,Giannetti2011,Johanning_2009,Giannetti2011,Schneider2012,Britton2012,PhysRevLett.111.147205,Islam2013,Jurcevic2014,Richerme2014,Bohnet2016,Keesling2019,2012.12268,PhysRevLett.124.063601}. These experimental techniques have also opened up the possibility to engineer an algebraically decaying many-body interacting potential, $1/R^{\alpha}$, as a function of the distance $R$, see Ref. \onlinecite{PhysRevA.99.012119}. However, most works in the condensed matter theory concern short-range (SR) interactions, mostly because in such SR systems analytical and numerical calculations are more tractable. However recently LR systems have been probed via powerfull novel numerical techniques such as tensor network approaches ~\cite{Vodola2015,LRBosonic,PhysRevA.96.043621,PhysRevB.97.155116,PhysRevLett.121.090603,PhysRevA.97.062301,PhysRevA.98.023607}, quantum Monte Carlo simulations~\cite{PhysRevE.68.056701,PhysRevB.93.104412, Humeniuk2020,2103.09469,2104.15070}, functional renormalization group\cite{PhysRevB.64.184106,PhysRevB.96.104432}, or high-order series expansions\cite{PhysRevE.92.022118,PhysRevLett.122.017203,PhysRevB.102.174424}. It turns out that such systems often display peculiar properties ~\cite{PhysRevB.100.144411,PhysRevLett.119.110601,PhysRevB.97.155113,PhysRevB.91.024415,Lepori2016,2009.04111,PhysRevLett.109.267203,PhysRevLett.111.207202,vanEnter2019,PhysRevLett.123.115701,Barma2019,1904.09937,Piccitto2019,PhysRevLett.124.063601,PhysRevB.100.014434,PhysRevResearch.2.013323,PhysRevLett.114.157201} that add odds with the standard folklore for SR systems. One of them is that even well away from a critical point the two-point correlation functions can have algebraically decaying tails\cite{Vodola2015,LRBosonic,PhysRevLett.119.110601,Chen2015,Chen2019}
thus apparently eliminating the concept of the diverging correlation length that is central to the theory of continuous phase transitions\cite{Sachdev}.

Depending on the exponent $\alpha$, the LR interactions in $D$ spatial dimensions can be classified into three regimes: weak decay of interactions for $\alpha<D$ (non-local regime), strong decay for $\alpha>D+1$ (local regime) and intermediate region $D<\alpha<D+1$ (weak non-local regime) \cite{PhysRevLett.29.917,PhysRevB.94.075156,PhysRevLett.121.240403,PhysRevB.96.104432,PhysRevA.98.023607,PhysRevLett.111.207202,PhysRevLett.111.260401}. As opposed to the generic Lieb-Robinson bound\cite{Lieb1972} in the SR models, the LR systems in the regime $\alpha>2D+1$ follow a generalized Lieb-Robinson bound defined with a generalized norm\cite{PhysRevLett.123.250605,PhysRevX.10.031009,PhysRevX.10.031010,2105.09960}. Although most of the LR systems are analytically intractable even in one dimension, there exists certain class of LR Hamiltonians that can be mapped into a quadratic Hamiltonian which is exactly solvable\cite{PhysRevLett.113.156402,Vodola2015,Maity2019}. In this case of non-interacting quasiparticles $\alpha>D+1$ is enough for the light-cone effect \cite{PhysRevX.10.031009}. 
One of such prototypical models is the long-range extended quantum Ising chain which is equivalent, via Jordan-Wigner transformation, to the long range Kitaev model. When $\alpha<1$, one has to consider a finite system in order for the thermodynamic limit to exist. Mean-field calculations are valid and the system effectively behaves like the Lipkin-Meshkov-Glick model~\cite{PhysRevLett.121.240403} with infinite range interactions. Also, the weak regime for $\alpha>2$ is not that interesting for our purpose because as model falls within the Ising universality class and behaves like the SR models\cite{PhysRevLett.113.156402,Vodola2015,LRBosonic}. Therefore, we focus here on the intermediate regime when $1<\alpha<2$. 

{\bf Ornstein-Zernike formula. ---}
Our starting point is the general form of the correlation function in SR systems: 
\be 
C_R \propto R^{-\eta} e^{-R/\xi},
\label{CRshort}
\ee
similar to the Ornstein-Zernike formula in Ising-like models. Here $\xi$ is the correlation length that depends on the distance from the critical point, $\epsilon$, like 
\be 
\xi\propto |\epsilon|^{-\nu}. 
\label{nu}
\ee
It discriminates between long- and short-range asymptotes of the correlation function. When $R\gg\xi$ then the tail of the correlation function can be approximated by an exponent, $C_R \propto e^{-R/\xi}$, in accordance with the folklore that away from the critical point correlations decay exponentially. However, the short-range asymptote,
\be 
C_R\propto R^{-\eta}
\label{eta}
\ee 
when $R\ll\xi$, demonstrates that the folklore is not quite correct. In fact this asymptote looks just like the correlation function at the critical point with the universal exponent $\eta$. In other words, even away from the critical point, up to the distance comparable with the correlation length, the correlations appear critical. We have to look at larger scale to notice that we are away from criticality. 

In the LR models the correlation function is the power law (\ref{eta}) at the critical point but, in distinction to the SR models, away from criticality the correlations also decay algebraically:
\be 
C_R\propto R^{-\tilde\eta}.
\label{tildeeta}
\ee 
There is no correlation length to be identified in this power law and, accordingly, the correlation length is not even mentioned in the LR models' literature.

{\bf Model. ---}
In order to proceed we adopt the exactly solvable long-range Kitaev chain \cite{Vodola2015,PhysRevLett.113.156402,Maity2019}
\bea
H_{K} &=& 2\mu \sum_n  
c_n^\dag c_n \nb 
- \sum_{n,r} J_r\left( c_n^\dag c_{n+r} + c_n^\dag c_{n+r}^\dag + {\rm H.c.} \right).
\label{HK}
\eea
Here $c_n$ are fermionic annihilation operators, $\mu=1-\epsilon$ is a chemical potential, and 
\be
 J_r = \frac{1}{\zeta (\alpha)} \frac{1}{r^\alpha}.
 \label{alpha}
\ee
is LR interaction depending on distance $r$ and normalized so that $\sum_r J_r =1$. Here $\zeta(\alpha)$ is the Riemann zeta function. After a Fourier transform the Hamiltonian becomes
\bea
H_{K} 
&=&
-2\sum_{k>0} 
\left( 1 - \epsilon -  {\rm Re}\tilde{J_k}\right) 
\left(c_k^\dag c_k+c_{-k}^\dag c_{-k}\right) + \nb \\
&& 
\sum_{k>0} 
{\rm Im}\tilde{J_k} 
\left(c_{k}^\dag c_{-k}^\dag + c_{-k}c_{k}\right).
\label{HkLR}
\eea
Here $\tilde{J_k}=\sum_r J_r e^{ikr}$ is a Fourier transform of $J_r$. For the considered $\alpha>1$ we have $\tilde{J_k}=\frac{{\bf Li}_{\alpha}(e^{ik})}{\zeta(\alpha)}$, where {\bf Li} is the polylogarithm function with ${\bf Li}_\alpha(x)=\sum_{n=1}^\infty\frac{x^n}{n^\alpha}$. The Hamiltonian is diagonalized by a Bogoliubov transformation
$
c_k ~=~ u_k  \gamma_k + v_{-k}^*  \gamma^\dagger_{-k}~,
$
where Bogoliubov coefficients satisfy stationary Bogoliubov-de Gennes equations:
\bea
\omega
\left(
\begin{array}{c}
u_k \\
v_k
\end{array}
\right)=
2\left[
\sigma^z\left( 1-\epsilon-{\rm Re}\tilde{J_k} \right) + \sigma^x{\rm Im}\tilde{J_k}
 \right]
\left(
\begin{array}{c}
u_k \\
v_k
\end{array}
\right)~~~~,
\label{BdGkLR}
\eea
with Pauli matrices $\sigma^{x,z}$. Their positive frequency eigenmodes yield quasiparticle spectrum
\be
\omega_k=
2
\sqrt{ \left( 1 - \epsilon - {\rm Re} \tilde{J_k} \right)^2 + \left( {\rm Im} \tilde{J_k} \right)^2 }.
\label{omegakLR}
\ee

{\bf Correlation length. ---}
As demonstrated in Appendix \ref{sec:asymp}, at the critical point $\epsilon=0$, the dispersion for small $k$ is   
$
\omega_k \propto k^{\alpha-1}
$, and
therefore the dynamical exponent $z=\alpha-1$. As a side comment, this $z<1$ imply that there is no speed limit to quasiparticle excitations explaining the absence of a sonic horizon \cite{PhysRevB.101.144429}. Moreover, away from the critical point,
$
\omega_k \propto |\epsilon|^1,
$
hence $z\nu=1$ and the correlation length exponent is $\nu=1/(\alpha-1)$. This way a length scale, that can be defined as
\begin{equation}
    \xi \propto |\epsilon|^{-1/(\alpha-1)},
    \label{xi}
\end{equation}
enters through the dispersion relation. By its very construction this length scale is relevant for the dispersion: $\xi^{-1}$ delimits the regime of small $k$. Therefore, we would expect that in real space $\xi$ demarcates the regime of large distances. We will see that indeed, via Bogoliubov eigenmodes, the length scale finds its way to correlations.

There are two quadratic correlators that fully characterize the Gaussian ground state:
\bea
A_R &= \langle c_{n+R}c_n^{\dagger}\rangle=&\frac{1}{\pi}\int_0^\pi|u_k|^2\cos{kR}~ dk     \label{alphaR}\\ 
B_R  &=\langle c_{n+R}c_n^{}\rangle=&\frac{1}{\pi}\int_0^\pi u_k v_k^*\sin{kR}~ dk. \label{betaR}
\eea
For large $R$ they both exhibit power laws that originate from non-analyticities of the Bogoliubov coefficients, $u_k$ and $v_k$, near $k=0$. 

\begin{figure}[t]
\includegraphics[width=\linewidth]{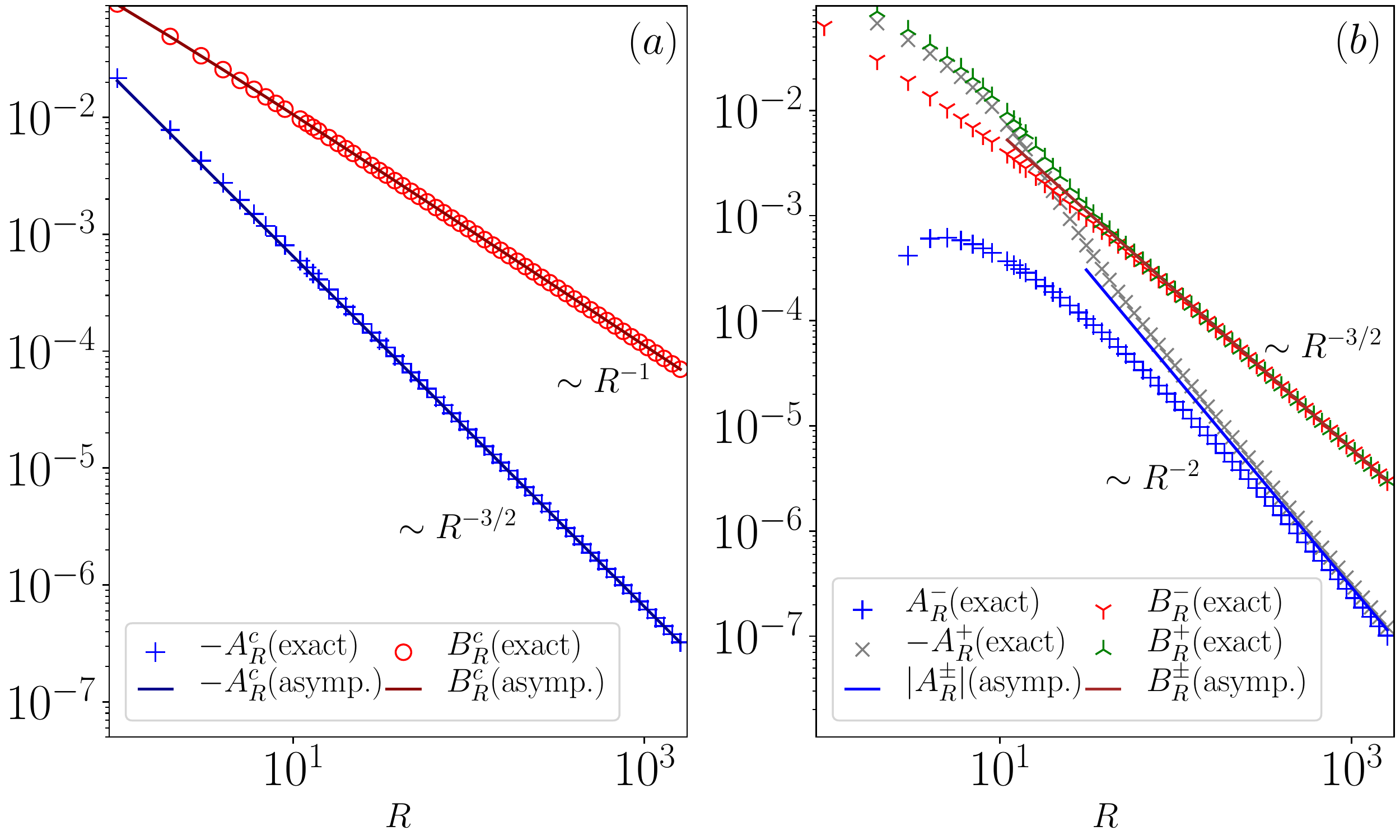}
\caption{
The correlation functions, $A_R$ and $B_R$, at the critical point (a) and and away from it (b). The data points were obtained with exact numerical integration while the solid lines are the asymptodes in, respectively, Eq. \ref{ABc} and Eqs. \eqref{AB+} and \eqref{AB-}. 
Here we set $\alpha=3/2$. In $(b)$, we have chosen $\epsilon=\pm0.5$ for ${A}_R^\pm$ and ${B}_R^\pm$.
}
\label{fig:alphabeta}
\end{figure}

{\bf At the critical point. -- } Using Eq. (\ref{asymtote}) we obtain asymptotes for small $k$:
\begin{align}
|u_k|^2   &\approx \Lambda^1_{U} \Big[1 + \Lambda^2_{U}  k^{2-\alpha}\Big], \nonumber \\
u_k v_k^* &\approx \Lambda^1_{V} [1 + \Lambda^2_{V} k^{2-\alpha}],
\label{eq:uvasymp}
\end{align}
where the coefficients are listed in Appendix \ref{sec:coef}. Inserting them in Eqs. (\ref{alphaR},\ref{betaR}) and extracting leading contributions to the integrals near $k=0$ we obtain the leading critical asymptotes for large $R$:
\begin{align}
 A_R^c \approx - \Lambda^c_A  \frac{1}{R^{3-\alpha}},~~~~
 B_R^c  \approx \frac {\Lambda^1_{V}}{\pi} \frac 1R.
 \label{ABc}
\end{align}
We can identify the respective exponents as
\be 
\eta_A=3-\alpha, ~~ \eta_B=1.
\label{etac}
\ee
In Fig. \ref{fig:alphabeta}(a) we compare the asymptotes with exact correlators. 


{\bf Negative ${\bf \epsilon}$. --} Using again Eq.~(\ref{asymtote}) we obtain asymptotes for small $k$:
\bea
&   |u_k|^2 & \approx 1 - \frac{{\rm Re}{(W_\alpha)}^2}{4|\epsilon|^2}|k|^{2(\alpha-1)},\nb \\
& u_k v_k^* & \approx \sign(k)\frac{{\rm Re}(W_\alpha)}{2|\epsilon|}|k|^{\alpha-1}  .
\label{UV_el0}
\eea
The minimal condition for these two asymptotes to be valid is that magnitude of the $k$- and $\epsilon$-dependent terms are much less than $1$. In both cases this means that $|k|^{\alpha-1}/|\epsilon|\ll1$. When Fourier-transformed it translates to length scales much longer than $\xi$ in Eq. (\ref{xi}). This demonstrates relevance of $\xi$ for correlations.

Inserting the asymptotes in Eqs. (\ref{alphaR},\ref{betaR}) and extracting leading contributions to the integrals near $k=0$, where $|k|^{\alpha-1}/|\epsilon|\ll1$, we obtain the leading off-critical asymptotes for large $R\gg\xi$:
\begin{align}
  A_R^- \approx \frac{\Lambda_A}{|\epsilon|^2}\frac{1}{R^{2\alpha-1}},~~~~
  B_R^- \approx \frac{\Lambda_B}{|\epsilon|} \frac{1}{R^\alpha},
  \label{AB-}
\end{align}
where the coefficients $\Lambda_{A,B}$ are listed in Appendix \ref{sec:coef}. We can identify the respective exponents as
\be 
\tilde\eta_A=2\alpha-1, ~~ \tilde\eta_B=\alpha.
\label{etaoff}
\ee
In Fig. \ref{fig:alphabeta}(b) we compare Eq. \eqref{AB-} with exact correlators. 


{\bf Positive ${\mathbf \epsilon}$. --} The asymptotes for small $k$ at positive $\epsilon$ can be expressed by the ones at negative $\epsilon$ as
\begin{equation}
    (u_k,v_k)_{\epsilon>0} = \sign(k) (v_k,u_k)_{\epsilon<0}
\end{equation}
which ultimately yields for positive $\epsilon$
\begin{align}
&   |u_k|^2  \approx \frac{{\rm Re}{(W_\alpha)}^2}{4|\epsilon|^2}|k|^{2(\alpha-1)}, \nb\\
&  u_k v_k^* \approx \sign(k)\frac{{\rm Re}(W_\alpha)}{2|\epsilon|}|k|^{\alpha-1}.
\end{align}
Inserting them in Eqs. (\ref{alphaR},\ref{betaR}) and extracting leading contributions to the integrals near $k=0$ we obtain the leading off-critical asymptotes for large $R$:
\begin{align}
A_R^+ \approx - \frac{\Lambda_A}{|\epsilon|^2}\frac{1}{R^{2\alpha-1}},~~~~
B_R^+ \approx  \frac{\Lambda_B}{|\epsilon|} \frac{1}{R^\alpha},
\label{AB+}
\end{align}
They are the same as \eqref{AB-} for negative $\epsilon$ except for the sign of $A_R$ and, again, their accuracy is limited to distances $R\gg\xi$.

\begin{figure}[t!]
\includegraphics[width=\linewidth]{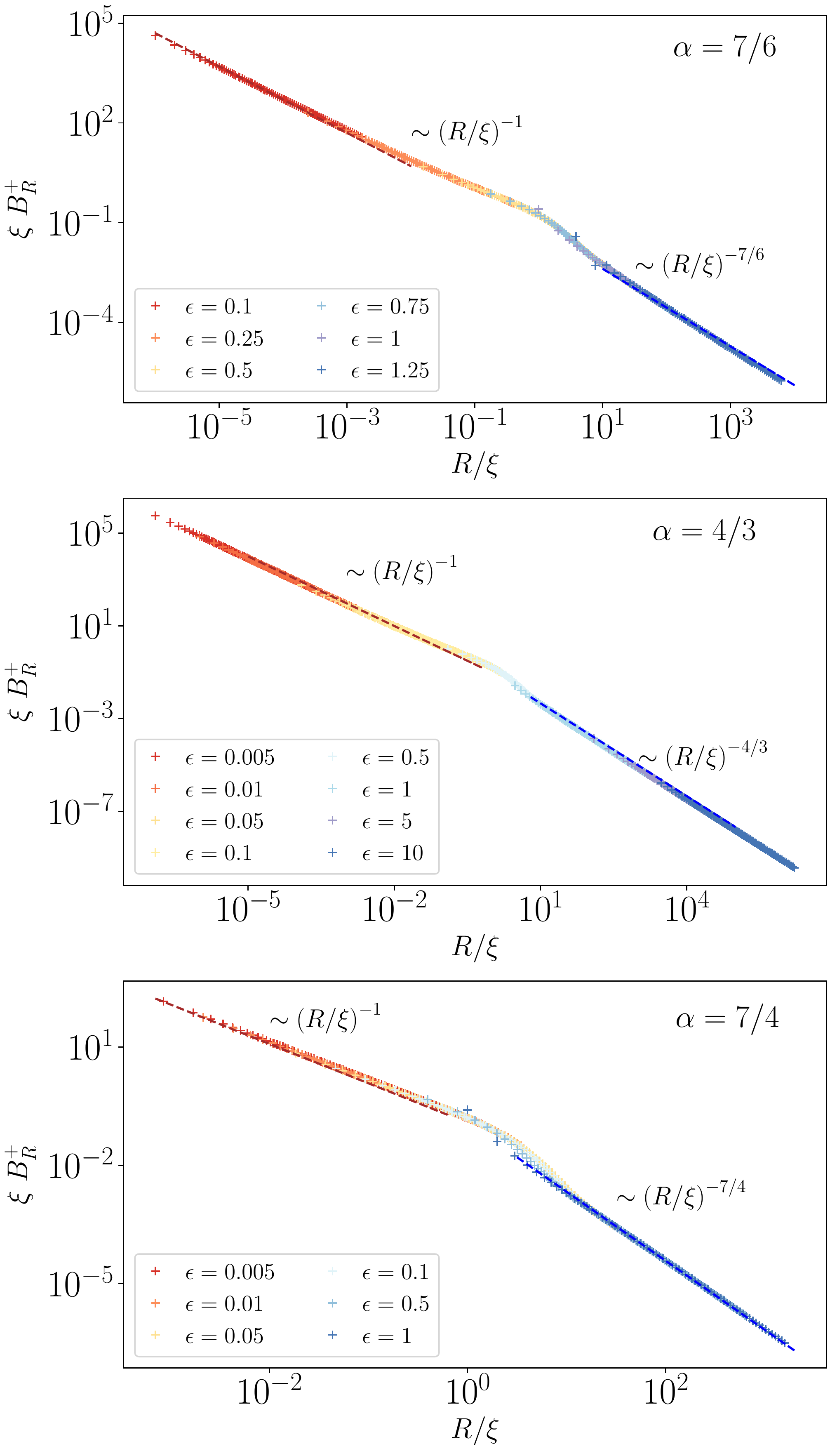}
\caption{
Plots of the scaled correlation function, $\xi^{\eta_B}B_R$, in function of scaled distance, $R/\xi$, for different values of the distance from the critical point, $\epsilon$. From top to bottom we show $\alpha=7/6$, $\alpha=4/3$, and $\alpha=7/4$, where $\alpha$ is the exponent of the long range interaction \eqref{alpha}. The collapse of the curves for different $\epsilon$ demonstrates that $\xi$ in \eqref{xi} is the only relevant scale of length, see the hypothesis \eqref{BSH}.
Here as in the following figures stability of numerical integration limits the distance to the range $1<R<1650$.
}
\label{fig:Bcollapse}
\end{figure}

{\bf Dominant/subdominant correlator. ---} 
After analyzing all the asymptotes, we can conclude that correlator $A_R$ is subdominant in the sense that for any considered $1<\alpha<2$ both the critical and the off-critical asymptotes of $A_R$ decay with $R$ faster, i.e. with a steeper exponent, than corresponding asymptotes of $B_R$:
\be 
\eta_A>\eta_B, ~~ \tilde\eta_A>\tilde\eta_B.
\ee 
It is interesting, and rather counter-intuitive, that its off-critical decay $A_R^{\pm}$ can be slower than its critical decay $A_R^c$:
\be 
\eta_A > \tilde \eta_A 
\ee 
when $\alpha<4/3$.

{\bf Scaling hypothesis. ---}
Each of the correlators $A_R$ and $B_R$ can be approximated by its critical asymptote in Eq. (\ref{ABc}) up to a certain $R$ where it begins to cross over to the long range power law tail in Eq. (\ref{AB-}) or (\ref{AB+}). In case of the dominant $B_R$ the crossover distance can be simply estimated as the $R$ where the two asymptotes are comparable, $B_R^c\approx B^\pm_R$:  
\begin{equation}
    R \propto |\epsilon|^{-1/(\alpha-1)} \propto \xi.
    \label{RB}
\end{equation}
This crossover length is proportional to the correlation length (\ref{xi}) that was inferred from the dispersion relation. 

This observation encourages us to formulate a scaling hypothesis for the dominant correlator:
\bea 
\xi^{\eta_B} ~B_R = F_B\left( R/\xi \right),
\label{BSH}
\eea 
where $\eta_B=1$ is the exponent of the critical correlator \eqref{ABc}, compare with its definition in Eq. \eqref{eta}. With $\xi=\epsilon^{-\nu}$ plots of the scaled correlator, $\xi^{\eta_B}B_R$, in function the scaled distance, $R/\xi$, for different $\epsilon$ should collapse to a common scaling function $F_B(x)$. We expect $F_B(x)$ to cross-over around $x\approx 1$ from $x^{-\eta_B}$ for small $x$ to $x^{-\tilde\eta_B}$ for large $x$. This is just what we can see in Fig. \ref{fig:Bcollapse}. The excellent collapse firmly establishes $\xi$ as the unique correlation length relevant for the dominant correlator. 

\begin{figure}[t]
\includegraphics[width=\linewidth]{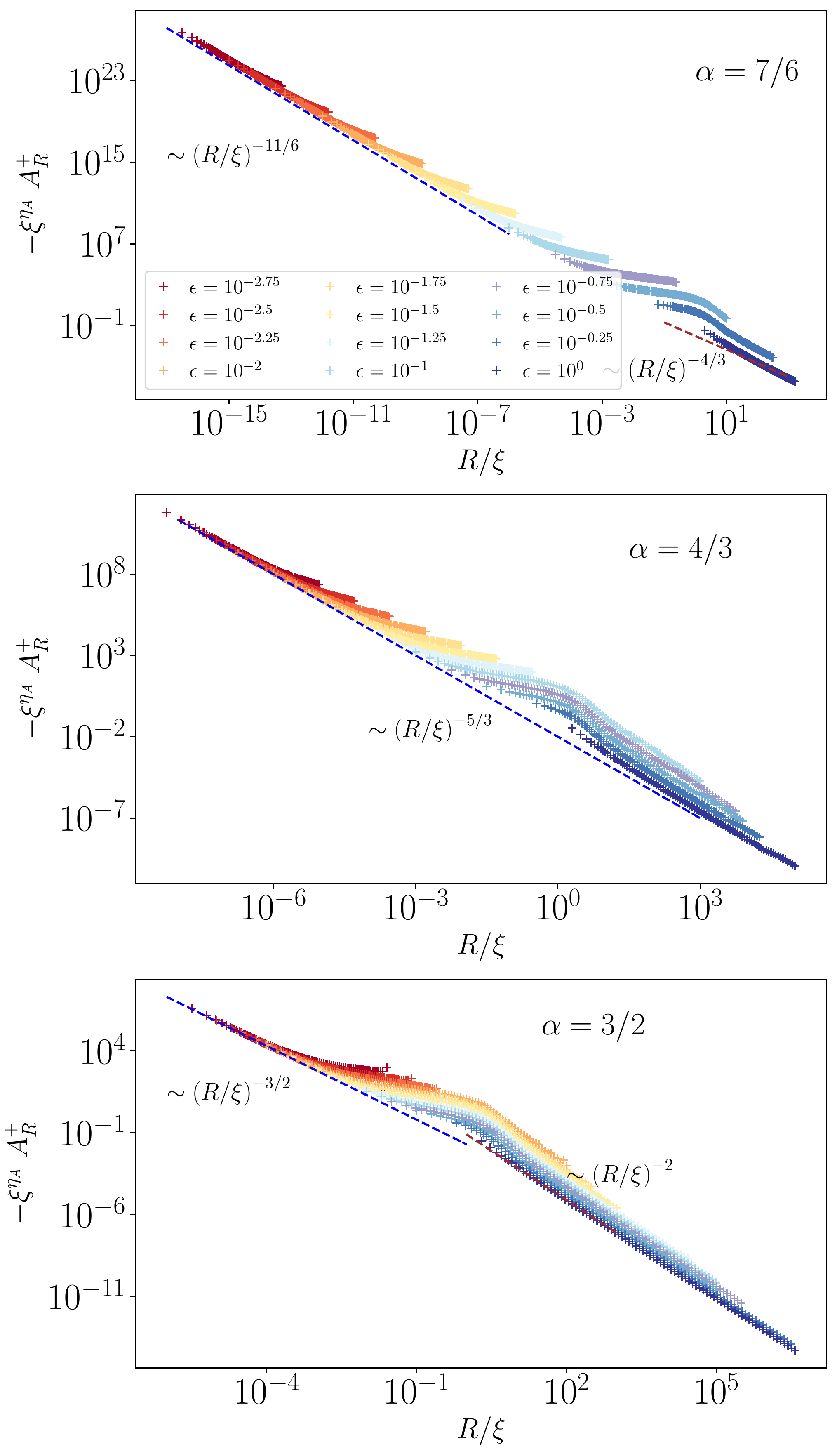}
\caption{
Plots of the scaled correlation function, $-\xi^{\eta_A}A_R^+$, in function of scaled distance, $R/\xi$, for different values of the distance from the critical point, $\epsilon$. From top to bottom, we show $\alpha=7/6$, $\alpha=4/3$, and $\alpha=3/2$, where $\alpha$ is the exponent of the long range interaction \eqref{alpha}. Here we used exponent $\nu=1/(\alpha-1)$ in \eqref{xi}. As predicted, for a given $\alpha$ the plots begin to follow the off-critical long-range asymptote \eqref{AB+} once $R/\xi$ becomes longer than $1$. In contrast, for small $R/\xi$ the curves approach the predicted critical slope $\eta_A$ in \eqref{ABc} but each of them reaches this asymptote for a different values of $R/\xi$. The short distance asymptote is reached at a second scale of length $\xi'\propto|\epsilon|^{\nu'}$ with a different exponent $\nu'$, see \eqref{xi'} and Fig. \ref{fig:Acollapse'}. 
}
\label{fig:Acollapse}
\end{figure}

\begin{figure}[t]
\includegraphics[width=\linewidth]{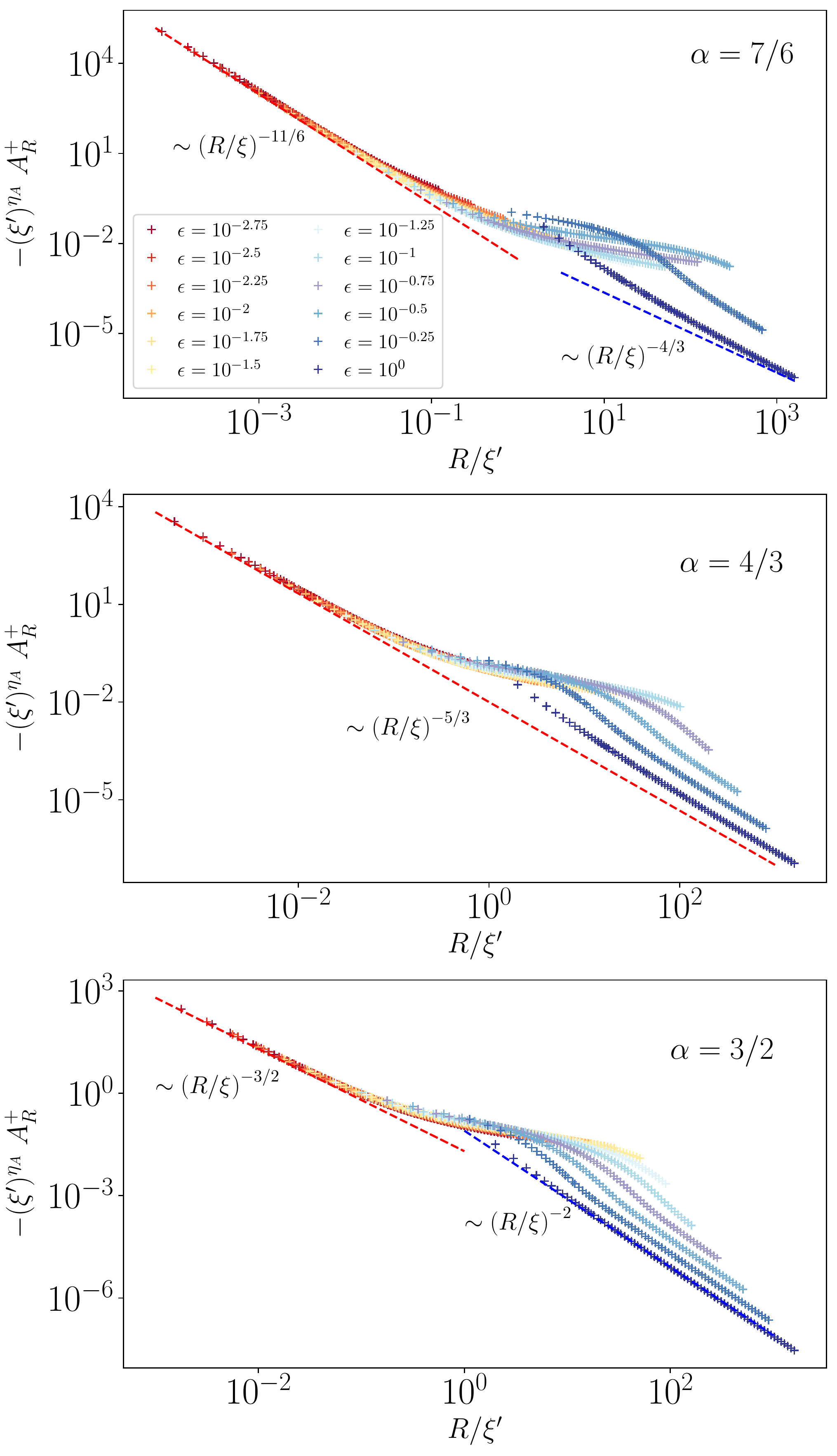}
\caption{
Plots of the scaled correlation function, $-(\xi')^{\eta_A}A_R^+$, in function of scaled distance, $R/\xi'$, for different values of the distance from the critical point, $\epsilon$. Here $\xi'$ is the second scale in \eqref{xi'}
with $\nu'=1/(\alpha-1/2)$. From top to bottom, we show $\alpha=7/6$, $\alpha=4/3$, and $\alpha=3/2$, where $\alpha$ is the exponent of the long range interaction \eqref{alpha}. The rescaling with $\xi'$ makes all the three panels collapse for $R/\xi'\ll1$. All the data are the same as in Fig. \ref{fig:Acollapse}, the only change is replacement of $\xi$ with $\xi'$. The collapse for $R/\xi'\ll1$ demonstrates that the critical asymptote \eqref{ABc} is valid up to $R\approx\xi'$.
}
\label{fig:Acollapse'}
\end{figure}

{\bf Subdominant correlator. ---} 
A simple scaling hypothesis as in Eq. \eqref{BSH} does not work for $A_R$, see Fig. \ref{fig:Acollapse}.
Therefore, $A_R$ must depend not only on $\xi$ but also on another second lengthscale $\xi'$. 
It can be inferred from, say, the asymptote in Eq. \eqref{AB+} 
where $A^+_R$ becomes ${\cal O}(1)$ when $R\propto|\epsilon|^{-1/{(\alpha-1/2)}}$. 
This defines the scale as
\be 
\xi'=|\epsilon|^{-1/{(\alpha-1/2)}}.
\label{xi'}
\ee 
For small enough $|\epsilon|$ we have $\xi'\ll\xi$ for any $1<\alpha<2$.
We demonstrate relevance of this $\xi'$ in Fig. \ref{fig:Acollapse'} 
where we plot scaled correlator, $\xi'^{\eta_A}A_R$, in function of scaled distance, $R/\xi'$. 
For each $\alpha$ the scaled plots with different $\epsilon$ collapse for $R/\xi'\ll1$ demonstrating relevance of $\xi'$
as a delimiter of small distances: $R\ll\xi'$. The crossover between short and long range correlations takes place between $\xi'$ and $\xi$. It can be formally encapsulated in a more general scaling hypothesis,
\be 
\xi'^{\eta_A} ~A_R = F_A\left( R/\xi' , \xi/\xi' \right),
\label{ASH'}
\ee 
with a scaling function of two arguments instead of just one. 

{\bf Conclusion. ---} 
The correlation length that can be formally identified in the dispersion relation for quasiparticle excitations also finds its manifestation in the dominant correlation function. As in the SR models, it marks a crossover between the short range critical power law (\ref{eta}) for $R\ll\xi$ and the long range asymptote (\ref{tildeeta}) for $R\gg\xi$. This asymptote is also a power law instead of the usual exponential decay with the correlation length $\xi$. Thus the correlation length plays a role in the correlation function but a more subtle one than for the SR models. In the same way as in the SR models, up to $R\approx\xi$ the correlations appear critical. The two asymptotes, (\ref{eta}) and (\ref{tildeeta}), constitute a generalized Ornstein-Zernike formula for long range systems and can be encapsulated in the scaling hypothesis \eqref{BSH}. 

Since the evaluation of the correlation length requires only the static preparation of a ground state, our hypothesis could easily be tested in cold-atomic experimental platforms\cite{iontrapexp1,Friedenauer2008,PhysRevLett.103.120502,Kim2010,PhysRevB.82.060412,Lanyon57,Giannetti2011,Johanning_2009,Giannetti2011,Schneider2012,Britton2012,PhysRevLett.111.147205,Islam2013,Jurcevic2014,Richerme2014,Bohnet2016,Keesling2019,2012.12268,PhysRevLett.124.063601}. In light of the surge of works related to LR models, we expect that this scaling hypothesis may be also useful in respect to thermal critical crossover\cite{PhysRevLett.119.110601, 2104.15070}, prethermalization\cite{Neyenhuis2017,1611.03992,2012.06505}, localization-delocalization transitions\cite{Smith2016,PhysRevResearch.3.013148} and dynamical quantum phase transitions\cite{PhysRevLett.119.080501,Zhang2017,PhysRevB.102.014303}.

\acknowledgments
This research was supported in part by the National Science Centre (NCN), Poland together with the European Union through QuantERA ERA NET program No.~2017/25/Z/ST2/03028 (DS) and by NCN under project 2019/35/B/ST3/01028 (JD).

\bibliographystyle{apsrev4-1}
\bibliography{static.bib}


\appendix

\section{Jordan Wigner, Fourier and Bogoliubov transformations}
\label{sec:JWetc}

After the Jordan-Wigner transformation,
\bea
&&
\sigma^x_n~=~
 -
 \left( c_n + c_n^\dagger \right)
 \prod_{m<n}(1-2 c^\dagger_m c_m)~, \\
&&
\sigma^y_n~=~
 i
 \left( c_n - c_n^\dagger \right)
 \prod_{m<n}(1-2 c^\dagger_m c_m)~, \\
&&
\sigma^z_n~=~1~-~2 c^\dagger_n  c_n~, 
\label{JordanWigner}
\eea
introducing fermionic operators $c_n$ that satisfy
$\left\{c_m,c_n^\dagger\right\}=\delta_{mn}$ and 
$\left\{ c_m, c_n \right\}=\left\{c_m^\dagger,c_n^\dagger \right\}=0$
the Hamiltonian $H$ becomes 
\be
 H_{K}~=~P^+~H^+_{K}~P^+~+~P^-~H^-_{K}~P^-~.
\label{Hc}
\ee
Above
$
P^{\pm}=\frac12\left[1\pm {\cal P} \right]
$
are projectors on subspaces with even ($+$) and odd ($-$) parity, 
\be 
{\cal P} ~=~
\prod_{n=1}^N\sigma^z_n ~=~
\prod_{n=1}^N\left(1-2c_n^\dagger c_n\right) ~
\label{P}
\ee
is the parity operator, and  
$
H^{\pm}
$
are corresponding reduced Hamiltonians. The $c_n$'s in $H^-$ satisfy periodic boundary condition, $c_{N+1}=c_1$, but the $c_n$'s in $H^+$ are anti-periodic: $c_{N+1}=-c_1$.

For definiteness, we can confine to the even parity. This actual choice makes no difference in the thermodynamic limit. The translationally invariant $H^+$ is diagonalised by a Fourier transform followed by a Bogoliubov transformation. The anti-periodic Fourier transform is  
\be
c_n~=~ 
\frac{e^{-i\pi/4}}{\sqrt{N}}
\sum_k c_k e^{ikn}~,
\label{Fourier}
\ee
where the pseudomomentum takes half-integer values
\be
k~=~
\pm \frac12 \frac{2\pi}{N},
\dots,
\pm \frac{N-1}{2} \frac{2\pi}{N}~.
\label{halfinteger}
\ee
Diagonalization of $H^+$ is completed by a Bogoliubov transformation
\be
c_k ~=~ u_k  \gamma_k + v_{-k}^*  \gamma^\dagger_{-k}~,
\label{Bog}
\ee
provided that Bogoliubov modes $(u_k,v_k)$ are eigenstates of stationary Bogoliubov-de Gennes equations with positive eigenfrequency $\omega_k$.

\section{Useful asymptotes}
\label{sec:asymp}

Using asymptotes of the polylogarithimic function \cite{NISThandbook}, we can expand  
\begin{align}
    {\bf Li}_\alpha(e^{ik}) = \Gamma(1-\alpha)(-ik)^{\alpha-1} + \sum_{n=0}^{\infty} \frac{\zeta(\alpha - n)}{n!}(ik)^n,
\end{align}
which ultimately gives
\begin{align}
    {\rm Re}(\tilde{J_k}) = {\rm Im}(W_\alpha)k^{\alpha-1} + \sum_{n=0}^{\infty} \frac{1}{n!}\frac{\zeta(\alpha-n)}{\zeta(\alpha)}\cos{(\frac{n\pi}{2})}k^n,  \\
   \intertext{and}
        {\rm Im}(\tilde{J_k}) = {\rm Re}(W_\alpha)k^{\alpha-1} + \sum_{n=0}^{\infty} \frac{1}{n!}\frac{\zeta(\alpha-n)}{\zeta(\alpha)}\sin{(\frac{n\pi}{2})}k^n.
\end{align}
Here $W_\alpha = \frac{\Gamma(1-\alpha)}{\zeta{(\alpha)}}e^{i\frac{\alpha\pi}{2}}$. When $k\to 0$ they simplify to
\bea
\lim_{k \to 0} {\rm Im}(\tilde{J_k}) &=& {\rm Re}{(W_\alpha)} k^{\alpha-1} + Z_\alpha k, \nonumber\\
\lim_{k\to 0} {\rm Re}(\tilde{J_k}) &=& {\rm Im}{(W_\alpha)} k^{\alpha-1} + 1, 
\label{asymtote}
\eea
with $Z_\alpha = {\zeta(\alpha-1)}/{\zeta(\alpha)}$. These asymptotic expressions are sufficient to obtain the quasiparticle spectrum for small $k$:
\bea
\omega_k \approx 
2 \Big[ \epsilon^2 + 2 \epsilon {\rm Im}(W_\alpha) k^{\alpha-1} + |W_\alpha|^2 k^{2(\alpha-1)}& \nonumber \\ 
+ 2 {\rm Re}(W_\alpha) Z_\alpha k^\alpha + Z_\alpha^2 k^2\Big]^{1/2}&  
 \label{omega_aymptote}
\eea

\section{Some coefficients}
\label{sec:coef}

\begin{align}
&\Lambda^1_{U} = {\frac{1}{2} \left( 1-\frac{{\rm Im}(W_\alpha)}{|W_\alpha|} \right)}, \\
&\Lambda^2_{U} = \frac{{\rm Re}(W_\alpha(\alpha))~{\rm Im}(W_\alpha(\alpha))~Z_\alpha(\alpha) }{|W_\alpha(\alpha)|^2~\big(|W_\alpha(\alpha)|-{\rm Im}(W_\alpha(\alpha))\big)}, \\
&\Lambda^1_{V} =  \frac{{\rm Re}(W_\alpha(\alpha))}{2|W_\alpha(\alpha)|}, \\ 
&\Lambda^2_{V} = \frac{Z_\alpha(\alpha)}{{\rm Re}(W_\alpha(\alpha))} -\frac{Z_\alpha(\alpha){\rm Re}(W_\alpha(\alpha))}{|W_\alpha(\alpha)|^2}, \\
&\Lambda^c_A = \frac{\Lambda_U^1 \Lambda_U^2}{\pi} {\Gamma(3-\alpha)\sin \frac{\alpha \pi}{2}}.
\end{align}

\begin{align}
& \Lambda_A =  -\frac{\Gamma(2\alpha)\sin{\alpha\pi}}{4\pi(2\alpha-1)}{\rm Im}(W_\alpha)^2, \\
& \Lambda_B = \frac{\Gamma(\alpha+2)\sin(\alpha\pi/2)}{2\pi \alpha(\alpha+1)}{\rm Re}(W_\alpha). 
\end{align}

\end{document}